\documentclass{llncs}

\usepackage{xspace,float,amsmath,url}

\usepackage{graphics}
\usepackage{epstopdf}
\usepackage{epsfig}

\usepackage{booktabs}

\newcommand{\exclude}[1]{}

\newcounter{lineno}

\newcommand{\utab}{\qquad}

\begin{document}

\title{Suffix arrays with a twist}

\author{Tomasz Kowalski$^\dag$, Szymon Grabowski$^\dag$, \\
Kimmo Fredriksson$^\ddag$ and Marcin Raniszewski$^\dag$}

\institute{
$^\dag$ 
  Lodz University of Technology, Institute of Applied Computer Science, \\
  Al.\ Politechniki 11, 90--924 {\L}\'od\'z, Poland \\
  \email{\{tkowals|sgrabow|mranisz\}@kis.p.lodz.pl} \\
$^\ddag$ 
  School of Computing, University of Eastern Finland, \\
  P.O.B. 1627, FI-70211 Kuopio, Finland, \email{kimmo.fredriksson@uef.fi} \\
}

\maketitle

\begin{abstract}
%%% The suffix array (SA) is a textbook full-text index, 
%%% popular for its simplicity and efficiency.
%The suffix array is a classic full-text indexing structure,
%combining effectiveness with simplicity.
%We present two classes of solutions attempting to improve
%efficiency of the suffix array even more, without augmenting
%it with extra structures (which would compromise the space usage).
%One embraces binary search alternatives, that is,
%non-standard traversal strategies during the pattern search.
%The other class of ideas concerns data layout,
%that is, non-standard permutations of suffix offsets.
%
The suffix array is a classic
%% , simple yet powerful, 
full-text index,
combining effectiveness with simplicity.
We discuss three approaches aiming to improve
its efficiency even more: changes to the navigation, 
data layout and adding extra data.
%% One concerns the navigation over the array,
%% the other data layout, and in the last one the suffix array
%% is augmented with extra data.
In short, we show that $(i)$ how we search for the right interval boundary
impacts significantly the overall search speed,
$(ii)$ a B-tree data layout easily wins over the standard one,
$(iii)$ the well-known idea of a lookup table for the
prefixes of the suffixes can be refined with using compression,
$(iv)$ caching prefixes of the suffixes in a helper array 
can pose a(nother) practical space-time tradeoff.
\end{abstract}

%%%%%%%%%%%%%%%%%%%%%%%%%%%%%%%%%%%%%%%%%%%%%%%%
\section{Introduction} \label{sec:intro}
%%%%%%%%%%%%%%%%%%%%%%%%%%%%%%%%%%%%%%%%%%%%%%%%
%% \noindent 
Everybody knows the suffix array (SA)~\cite{MM90}, a simple full-text index
data structure capable of finding the $occ$ occurrences of a pattern $P$
of length $m$ in $O(m\log n + occ)$ time, where $n$ is the 
length of the indexed text.
The search mechanism consists in two binary searches, 
for the left and the right boundary of the interval of text suffixes 
starting with $P$, in the array of suffix offsets arranged in the 
lexicographical order of their text content.
The performance of the suffix array can serve as a measuring stick 
for more advanced (e.g., compressed) text indexes~\cite{FGNVjea08} 
and at least for this reason it is important to know how to implement 
it efficiently and what space-time tradeoffs are possible.

%The suffix array can be perceived as a simplification 
%of the suffix tree (ST)~\cite{Wei73}, 
%a trie whose string collection is the set of all the 
%suffixes of a given text, with an additional requirement that all 
%non-branching paths of edges are converted into single edges. 
%Indeed, a suffix array can be obtained from a suffix tree 
%by visiting its leaves in order (from left to right, obtained 
%by depth-first traversal of the ST).
%Depending on the implementation, the pattern search over ST takes 
%either $O(m\log\sigma + occ)$ or $O(m + occ)$ time (where the latter variant 
%involves perfect hashing).
%The common wisdom about the practical performance of ST and SA 
%is that they are comparable, but Grimsmo~\cite{Gri07} showed that 
%a careful ST implementation 
%may be up to about 50\% faster than SA if the number of matches is very small 
%(in particular, one hit), but if the number of hits grows, the SA becomes 
%more competitive, sometimes being an order of magnitude faster.

The body of research on engineering the suffix array is 
%% however 
surprisingly scarce.
Although the basic SA idea can be easily grasped even by high-school students, 
many design choices from the implementor's point of view are not obvious.
Let us pose a few questions:
$(i)$ Can the binary search strategy be replaced with a faster one, 
e.g., based on interpolation search?
$(ii)$ As $occ$ is usually small, what is practically the best way 
to find the right interval boundary once the left boundary is known? 
$(iii)$ Can we change the data layout of suffixes in order to obtain 
more local memory accesses? 
%% $(iv)$ Can the SIMD instructions offered by most modern CPUs speed up the search? 
$(iv)$ How can we augment the suffix array with a moderate amount of extra data, 
to initially reduce the search interval and/or speed up string comparisons?
The answer to some of them is known,
%% (e.g., already in the seminal 
%% Manber and Myers paper~\cite{MM90} an answer to the last of the 
%% listed questions was given, with the idea of precomputing SA intervals 
%% for all possible $k$-long prefixes), 
yet in this work we are dealing with 
the mentioned issues in a more systematic way.

%% We use standard notation. 
%% Let $T[1 \ldots n]$ be a text of length $n$ 
%% and $P[1 \ldots m]$ a pattern of length $m$.
%% Both $T$ and $P$ are composed of symbols from an integer alphabet 
%% $\Sigma = \{1, \ldots, \sigma\}$.

%%%%%%%%%%%%%%%%%%%%%%%%%%%%%%%%%%%%%%%%%%%%%%%%
\section{Ideas and incarnations}
%%%%%%%%%%%%%%%%%%%%%%%%%%%%%%%%%%%%%%%%%%%%%%%%
%% \noindent 
The considered ideas are divided into three groups and each of them 
is described in a separate paragraph.
First we discuss non-standard SA traversal strategies.
Later we advocate for alternative data layouts, beneficial for the search speed.
In the last paragraph some ways to augment the suffix array 
with extra data, to make the pattern search even faster, are proposed.

%%%
%% \subsection{Navigating over the suffix array}
\paragraph*{\textbf{Navigating over the suffix array.}}
%
%Wide registers together with single-instruction multiple-data (SIMD) 
%instruction sets are a standard extension of modern CPUs, 
%including Intel's Pentium 4, Core 2, Nehalem and more recent architectures, 
%Intel's Xeon, AMD's Phenom and Bulldozer, and ARM Cortex-A mobile processors.
%
%Zhou and Ross~\cite{zhou2002implementing} 
%proposed a SIMD-ized version of binary
%search that is geared towards small datasets (up to a few hundred keys). 
%The key idea of their approach is to replace
%the comparison of just the separator with the search key by ...

A textbook alternative to binary search is interpolation search, 
which performs a number of ``guesses'' concerning the query's location 
based on the query value and the assumed distribution of keys.
It is well-known that interpolation search over constant-size keys achieves 
$O(\log\log n)$ expected time not only for the simplest case, i.e., uniformly 
random distribution~\cite{Willard85a}, yet we are not aware of any 
published experiments regarding text suffixes.
Unfortunately, a straightforward interpretation of string prefixes
(which have a lot of duplicates) as integers and standard linear 
interpolation yielded rather disappointing preliminary results. 
%% More work is needed in order to close the gap to $O(\log\log n)$ search steps.

Another question concerning the navigation over the SA is how the right interval 
boundary should be found.
We examine two methods: a naive one performs the binary search over 
the range $left \ldots n$ of suffixes, where $left$ is the position of the 
least suffix greater or equal to the pattern, 
and the doubling (galloping) algorithm, which peeks the locations
%% $SA[left+1], SA[left+2], SA[left+4], \ldots$, until $SA[left + 2^i]$ 
$SA[left + 2^i], i = 0, 1, 2, \ldots$, until it reaches too far and 
the search continues in the binary manner over the last considered interval.
Note that the time complexity of the right interval boundary search improves 
in this way from $O(m\log n)$ to $O(m\log occ)$.

%%%
%% \subsection{Linearized $k$-ary tree data layout}
\paragraph*{\textbf{Linearized $k$-ary tree data layout.}}
Binary search over a sorted array is equivalent to walking down a path 
in a complete binary search tree. 
Schleger et al.~\cite{schlegel2009k} noticed that changing the 
%% sorted array 
tree 
layout 
from binary to $k$-ary ($k > 2$), together with linearization of the search tree, 
may be more cache-friendly and also convenient for SIMD processing.
%% In their experiments, on arrays of integers or floats 
%% and an Intel Core i7 CPU, 
%% it achieved a speedup of as much as 3 up to 4.5 for 32-bit data 
In their experiments 
%% on an Intel Core i7 CPU 
(Intel Core i7) 
it achieved a speedup of as much as 3 up to 4.5 for 32-bit numbers 
and 2 to 2.5 for 64-bit 
%% keys,
numbers,
compared to a plain binary search.
This data organization can also be called an (implicit) B-tree layout~\cite{KM15}.
%% where the case of $B = 1$ (a complete binary tree with the root going first, 
%% then followed by its both children, etc.) is called the Eytzinger layout 
%% or the heap-order layout.
We apply the presented idea to the suffix array, which, to our knowledge, 
has not been tried before.
Note that setting the B-tree layout for a suffix array cannot be 
comparably successful as for, e.g., integers, as the accesses to the text 
are still at ``random'' areas.
%% On the other hand, 
%% We mitigate this issue with storing extra data, 
%% as described by the end of the next paragraph.

%%%
%%\subsection{Augmenting the suffix array}
\paragraph*{\textbf{Augmenting the suffix array.}}
Manber and Myers in their seminal paper~\cite{MM90} presented a nice 
trick saving several first steps in the binary search:
if we know the SA intervals for all the possible first $k$ symbols of the
pattern, we can immediately start the binary search in a corresponding interval.
We can set $k$ to $\log_{\sigma} n$, where $\sigma$ is the alphabet size, 
with $O(n\log n)$ extra bits of
space and constant expected size of the interval.
%% , which leads to $O(m)$ average search time.
Unfortunately, real texts are far from random, hence in practice, 
%% if text symbols are bytes, 
we can use $k$ up to 3 (assuming that text symbols are bytes), 
which offers a limited (yet, non-negligible) benefit. 
This idea will be referred in our experiments as using a lookup table, 
and more specifically we will denote the lookup table on pairs (resp. triples) 
of symbols with LUT2 (resp. LUT3).

In the same spirit, Grabowski and Raniszewski~\cite{GR14} 
%% recently proposed storing the intervals for all $k$-symbol strings 
%% {\em occurring in the text} in a hash table.
use a hash table to store the intervals for all $k$-symbol strings 
{\em occurring in the text}.
This can significantly reduce the initial interval for real texts with 
relatively little extra space.

%% In their experiments $k$ was set to 8 (for most datasets) and xxhash (\url{https://code.google.com/p/xxhash/}) 
%% used as the hash function.

In this work we first propose a lookup table with keys being concatenations 
of Huffman codewords for the starting symbols of the text suffixes 
(Table~\ref{table:luthuff}), truncated to a specified length of $b$ bits.
Pattern search translates to finding the first $b$ bits of Huffman encoding 
of the pattern, which is the LUT key, 
and then following with binary search over a range of suffixes 
read from the LUT.
A correct implementation of this idea requires a reordering of the suffixes 
in the SA, to avoid nested LUT ranges 
(other options, like replacing Huffman with Hu--Tucker coding, are also 
possible but we have not tried them out).
Note also that the Huffman-based LUT entries store twice more data 
(both boundaries of the interval) than in the standard LUTs. 
%% , hence e.g. LUT-huff-15b is equivalent in space to LUT2.

\begin{table}
\centering
\caption{Average binary logarithms of the search interval widths for different LUT variants 
(first three rows: Huffman encoding with 16--24 bits, next two rows: standard 2-/3-byte LUTs).
%% The first three rows correspond to Huffman encoding and 16, 20 or 24 bits. 
%% The next two rows are for the standard LUTs over 2 or 3 bytes. 
The averages are over 100K patterns taken randomly from the text.}
\label{table:luthuff}
\setlength{\tabcolsep}{0.45em}
%% \begin{tabular}{L{2.45cm} R{1.85cm} C{1.5cm} C{1.5cm}}
\begin{tabular}{lrrrrr}
\toprule
dataset        & space (MiB) & ~~~~~dna200  & ~english200  & proteins200 & ~~~~~xml200  \\
\midrule
---            & ---         & 27.644  & 27.644      & 27.644      & 27.644 \\
LUT-Huff-15b   & 0.25        & 14.453  & 15.530      & 12.858      & 17.045 \\
LUT-Huff-19b   & 4.00        & 11.070  & 13.432      &  9.103      & 15.898 \\
LUT-Huff-23b   & 64.00       &  7.790  & 11.630      &  5.617      & 14.985 \\
LUT2           & 0.25        & 23.738  & 19.502      & 19.266      & 18.910 \\
LUT3           & 64.00       & 21.820  & 16.555      & 15.108      & 16.756 \\
\bottomrule
\end{tabular}
\end{table}

We also propose mixing the LUT or hash table interval narrowing with 
the B-tree layout, and also augmenting the search tree with 
prefixes of the suffixes in several top levels of the B-tree. 
Copying these text snippets into a helper array is beneficial due to 
more local memory accesses.

%%%%%%%%%%%%%%%%%%%%%%%%%%%%%%%%%%%%%%%%%%%%%%%%
\section{Experimental results}
%%%%%%%%%%%%%%%%%%%%%%%%%%%%%%%%%%%%%%%%%%%%%%%%

%% All experiments were run on a machine equipped with 
The test machine 
%% sported 
was equipped with 
a 4-core Intel i7 4790 3.6\,GHz CPU and 32\,GB of 1600\,MHz DDR3 RAM (9-9-9-24), 
hosting Windows Server 2012 R2.
%% The CPU cache sizes were:
%% $4 \times 32$\,KB (data) and $4 \times 32$\,KB (instructions) in L1, 
%% $4 \times 256$\,KB in L2 
%% and 8\,MB in L3 level.
%% One CPU core was used for the computations.
All codes (\url{https://bitbucket.org/kowallus/sa-search-dev/})
were written in C++ and compiled with 64-bit gcc 4.9.3 
with \texttt{-O3}.
%% \texttt{-O3} and \texttt{-mpopcnt} options.
%% Our source codes are available at
%% \url{https://bitbucket.org/kowallus/sa-search-dev/}.
All presented times are averages over 500K patterns taken randomly 
from the text.
The searches were performed over 
200\,MB datasets from the well-known Pizza~\&~Chili corpus.
%% were used as the texts.
%% \footnote{\url{http://pizzachili.dcc.uchile.cl/}}.

\begin{figure}
\centerline{
\includegraphics[width=0.45\textwidth,scale=1.0]{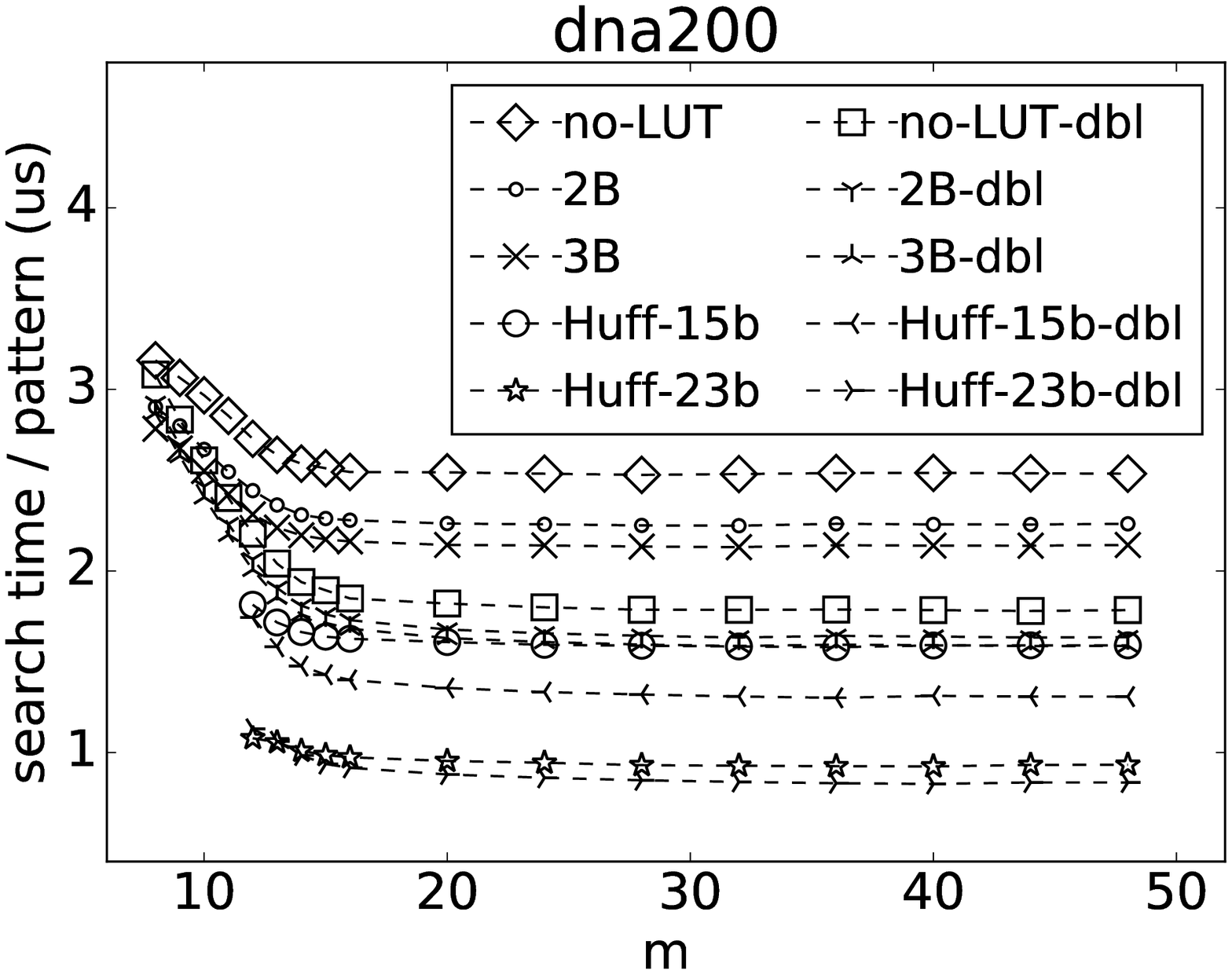}
\includegraphics[width=0.45\textwidth,scale=1.0]{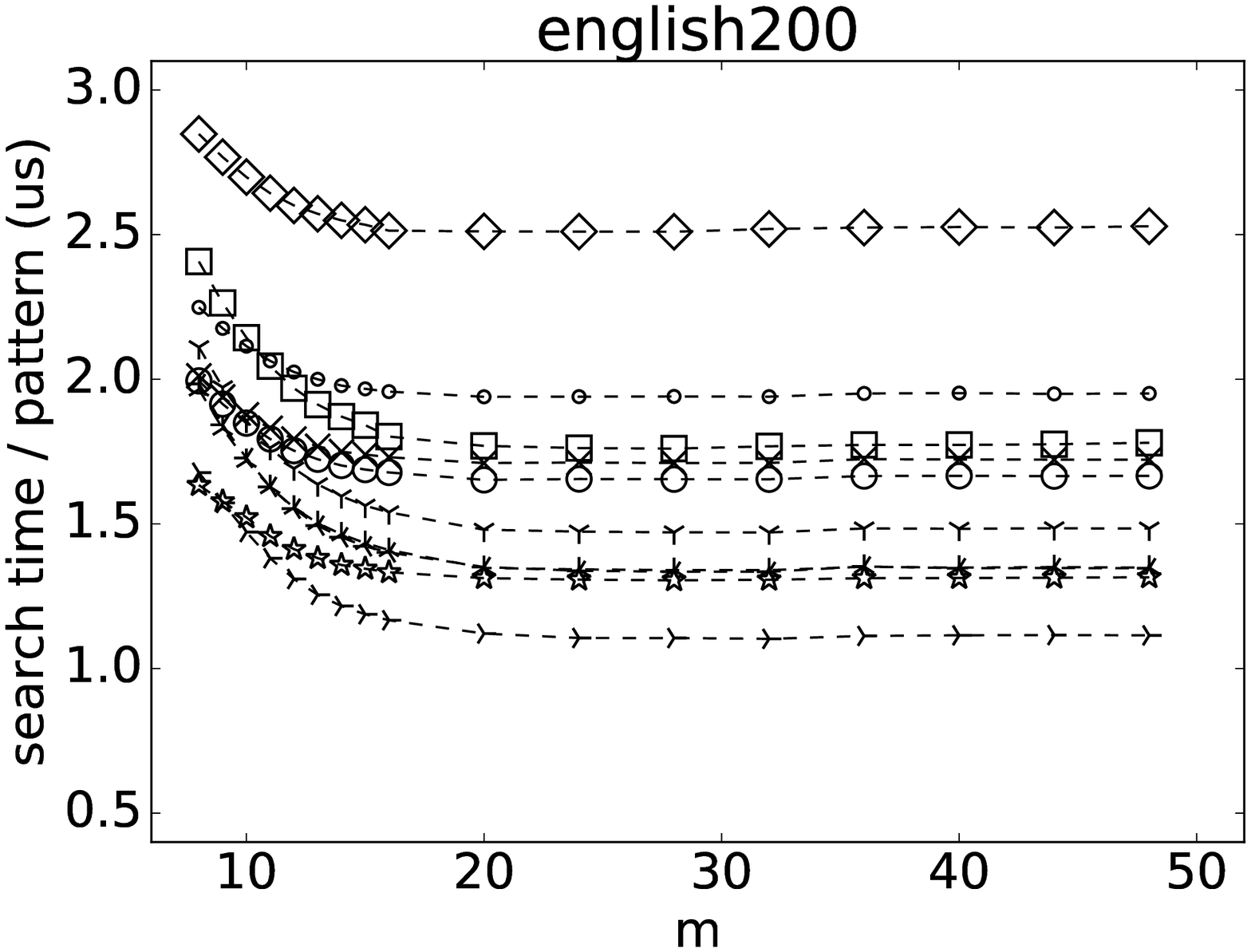}
}
\centerline{
\includegraphics[width=0.45\textwidth,scale=1.0]{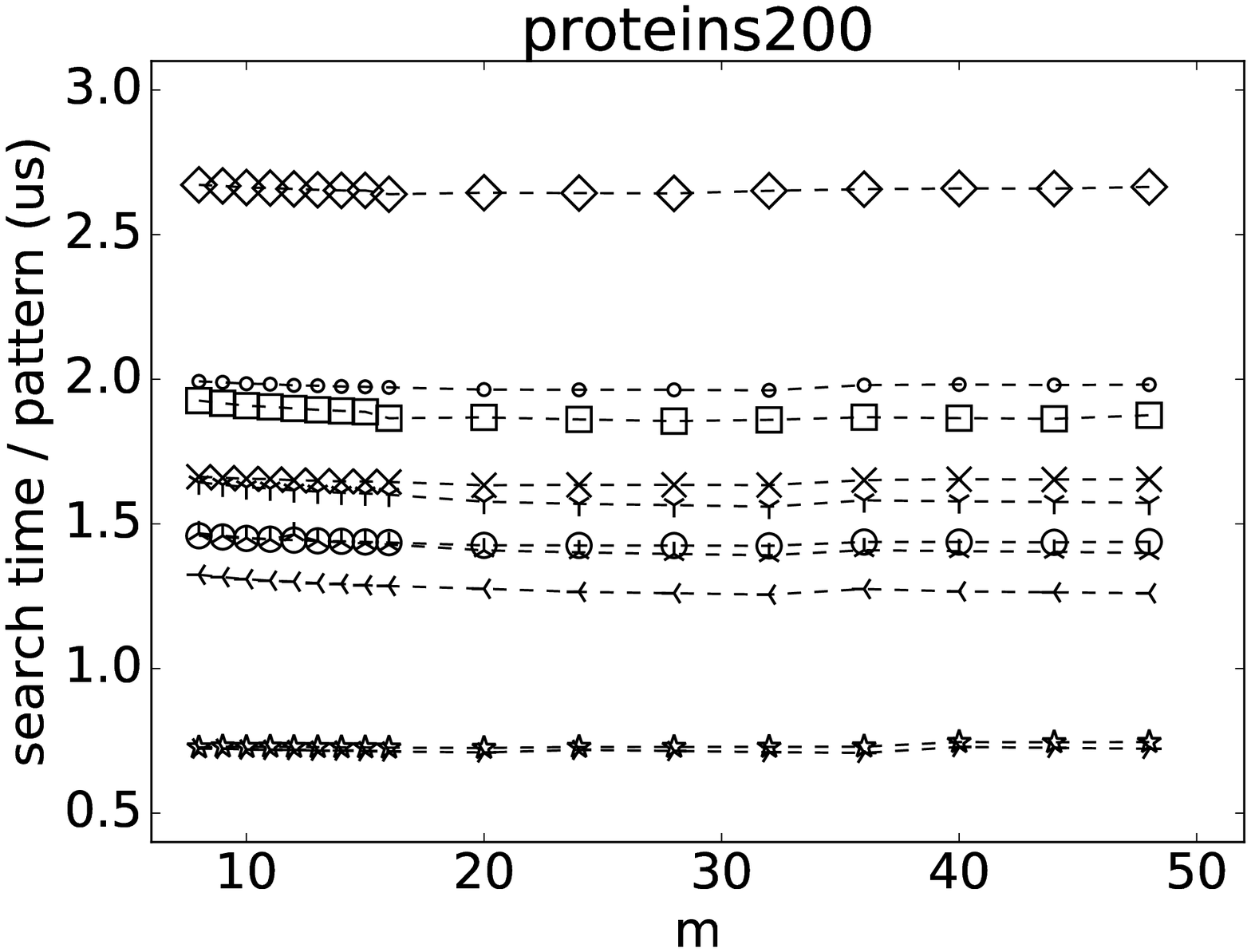}
\includegraphics[width=0.45\textwidth,scale=1.0]{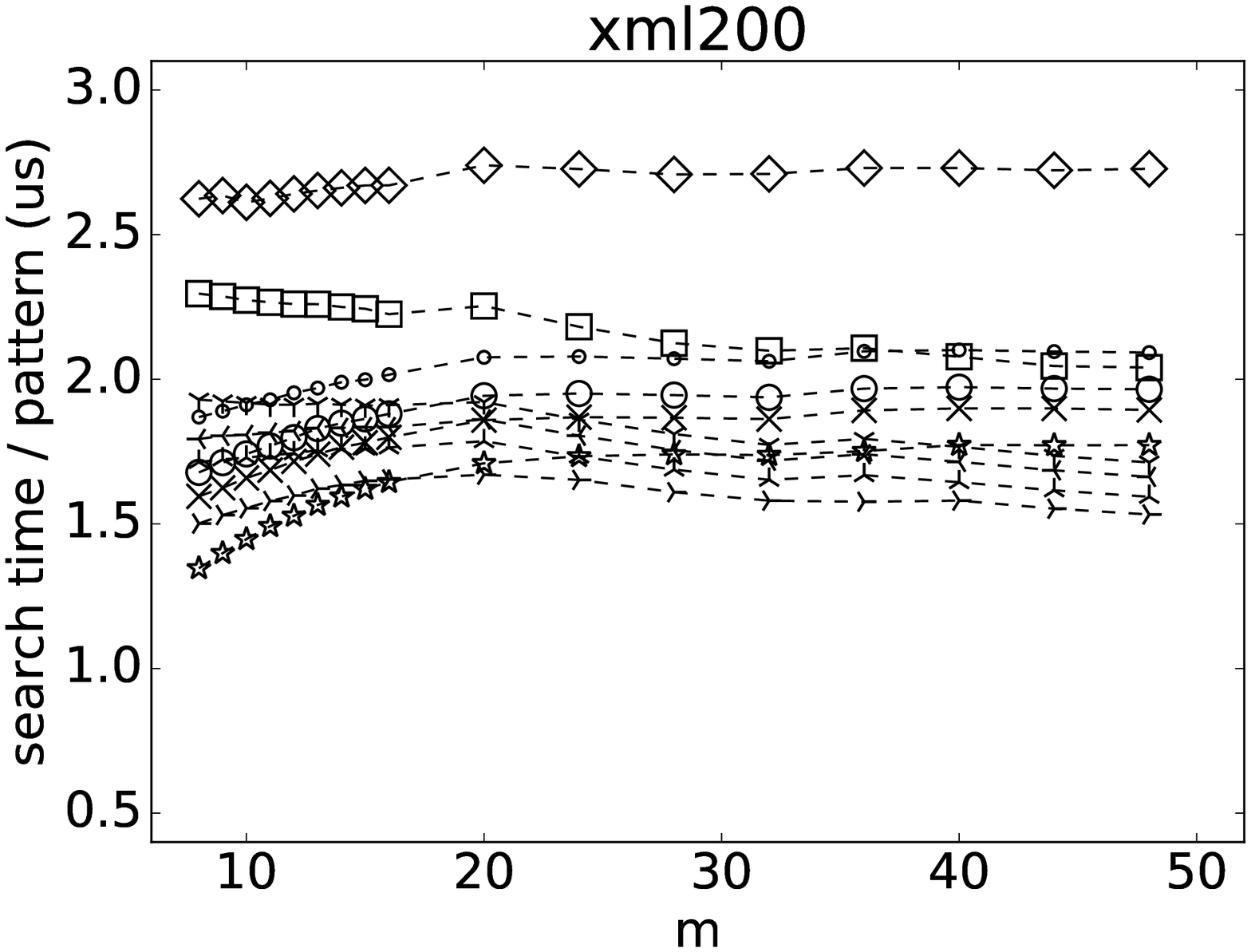}
}
%%\centerline{
%%\includegraphics[width=0.45\textwidth,scale=1.0]{xml200_count.eps}
%%}
\caption[Results]
{Count times for the standard suffix array and the SA augmented with 
%% a lookup table on pairs (2B) or triples of bytes (3B), 
%% and on 15 (Huff-15b) or 23 (Huff-23b) bits of Huffman codewords,
a lookup table on pairs or triples of bytes,
and on 15 or 23 bits of Huffman codewords,
using the standard and the doubling 
search for finding the right interval boundary.}
\label{fig:sa_rightboundary}
\end{figure}

In the first experiment we show how the count times are affected by
two things: 
using lookup tables, including the introduced Huffman-based ones 
(on 15 or 23 bits)
and choosing a proper interval's right boundary search 
(Fig.~\ref{fig:sa_rightboundary}). 
The doubling trick reduces the times usually by 20--30\% 
for the standard and LUT2-boosted suffix array,
yet the effect is smaller for short patterns, especially for DNA 
(where short patterns tend to have thousands of occurrences).
This can be explained by the relatively small difference between 
$\log n$ and $\log occ$ in those cases.
The Huffman-based LUTs are more efficient than their traditional 
counterparts 
(when about the same amount of memory is sacrificed).
%% (using the same amounts of memory).

\begin{figure}
\centerline{
\includegraphics[width=0.42\textwidth,scale=1.0]{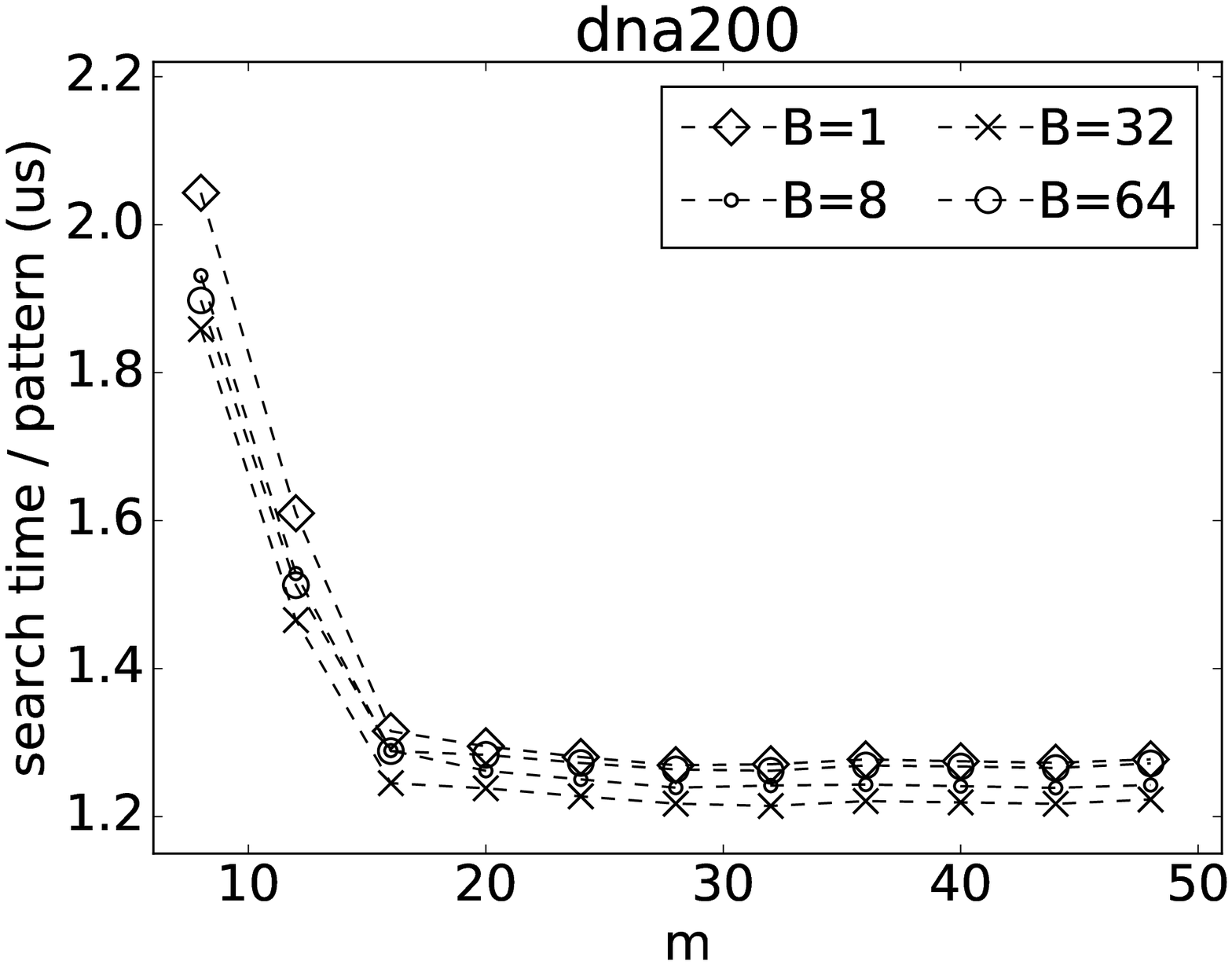}
\includegraphics[width=0.42\textwidth,scale=1.0]{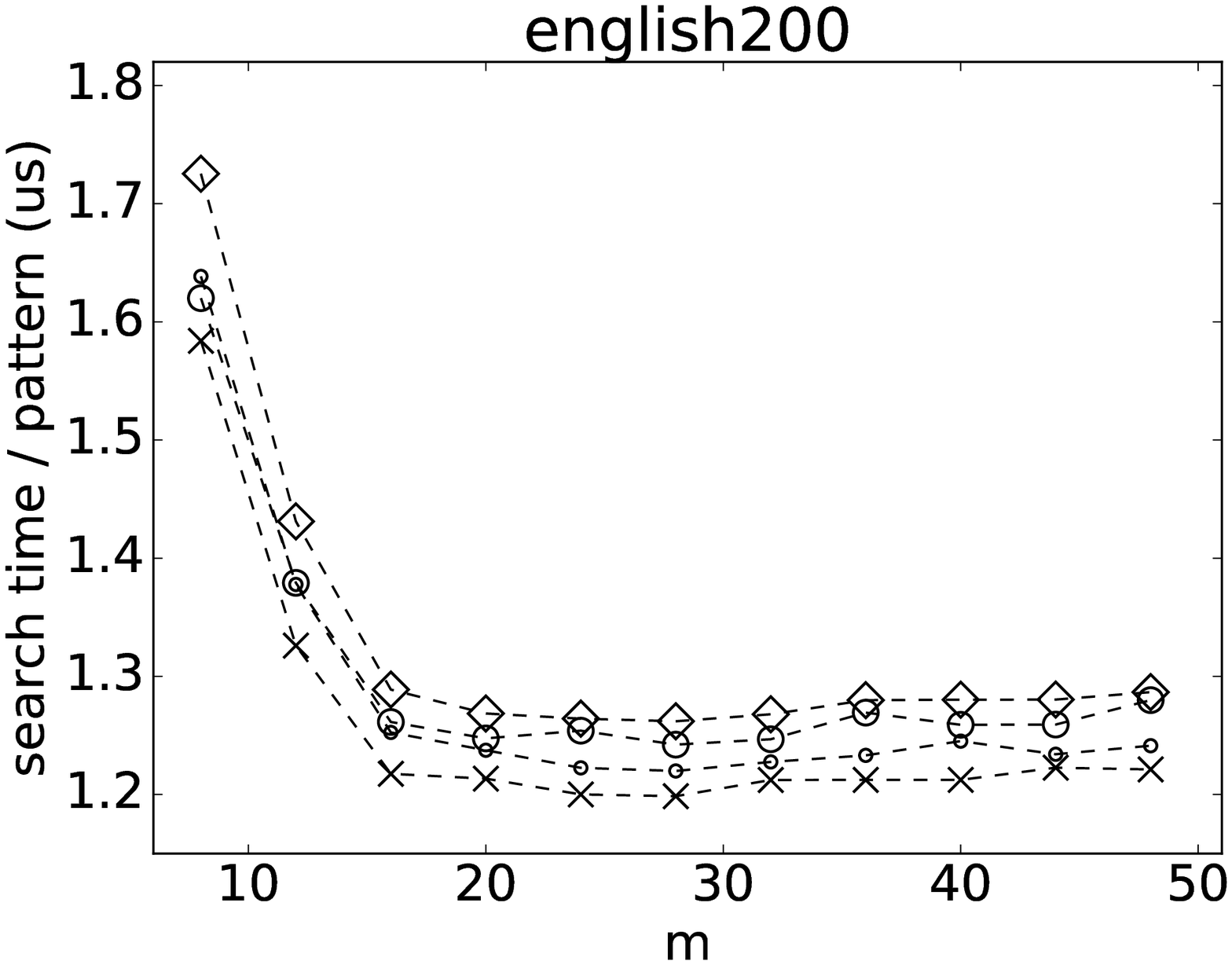}
}
\centerline{
\includegraphics[width=0.42\textwidth,scale=1.0]{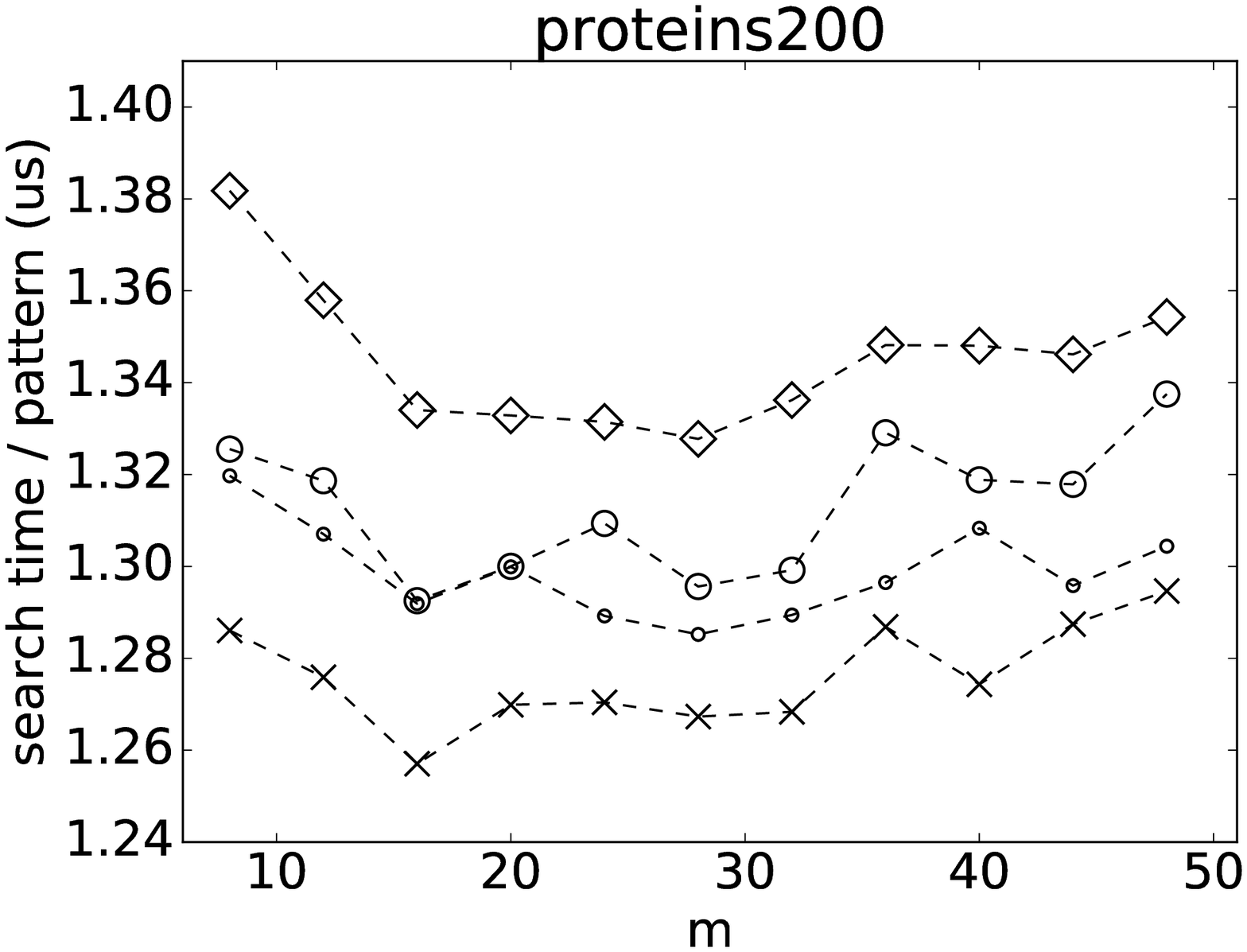}
\includegraphics[width=0.42\textwidth,scale=1.0]{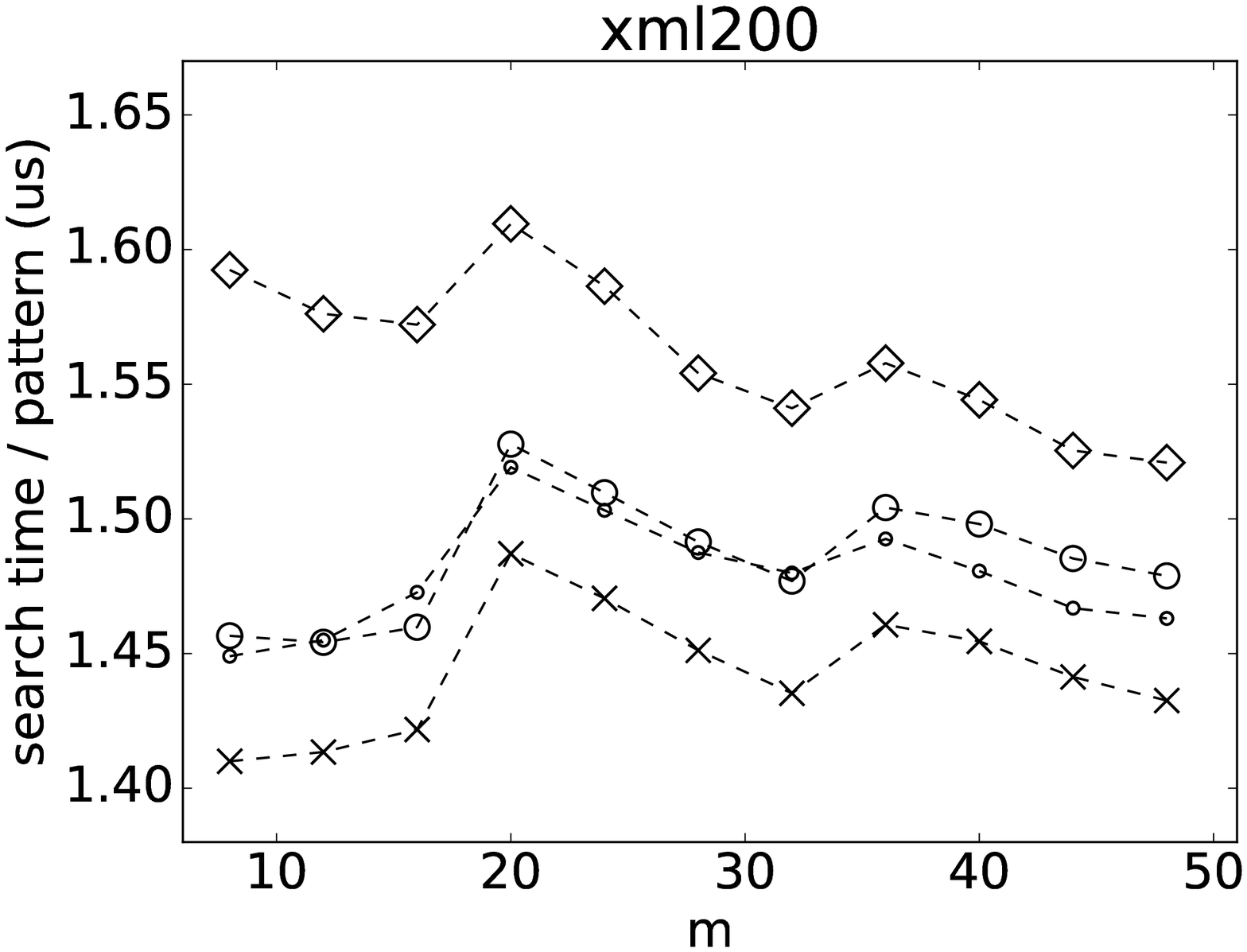}
}
\caption[Results]
{Count times for the SA with the B-tree layout, for selected node sizes $B$.}
\label{fig:sa_kary}
\end{figure}

Fig.~\ref{fig:sa_kary} shows the impact of the node size $B$ in the 
B-tree layout on the count times, with varying pattern length.
Even $B=1$ results in a much faster search than with a standard SA 
(by a factor of 1.7--2.0; cf.~also Fig.~\ref{fig:sa_rightboundary}) 
and growing $B$ helps more, 
up to $B = 32$ (on all the datasets, $B=64$ is slightly slower).
Still, the speed gap between $B=1$ and $B=32$ rarely exceeds 10\%.

\begin{figure}
\centerline{
\includegraphics[width=0.42\textwidth,scale=1.0]{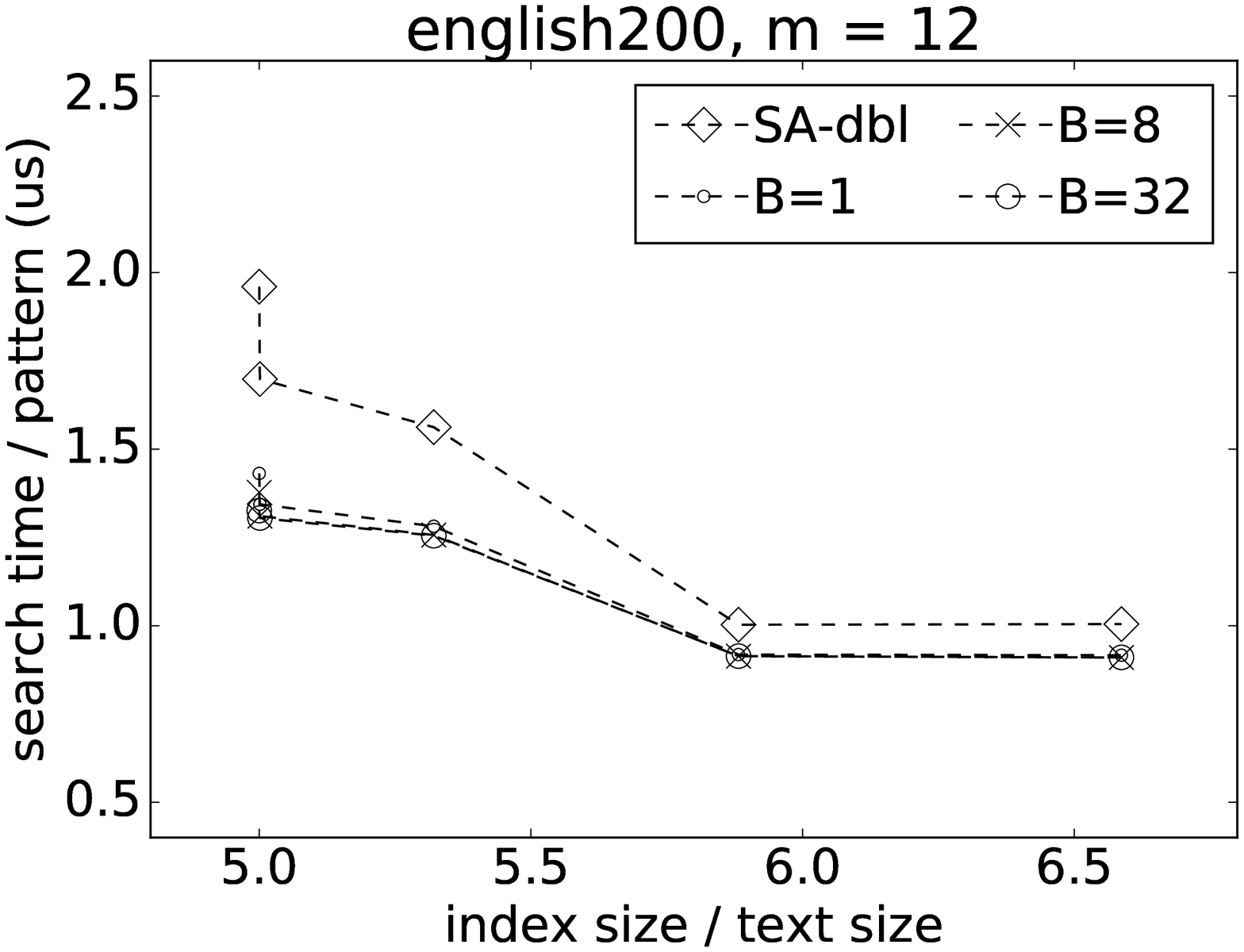}
\includegraphics[width=0.42\textwidth,scale=1.0]{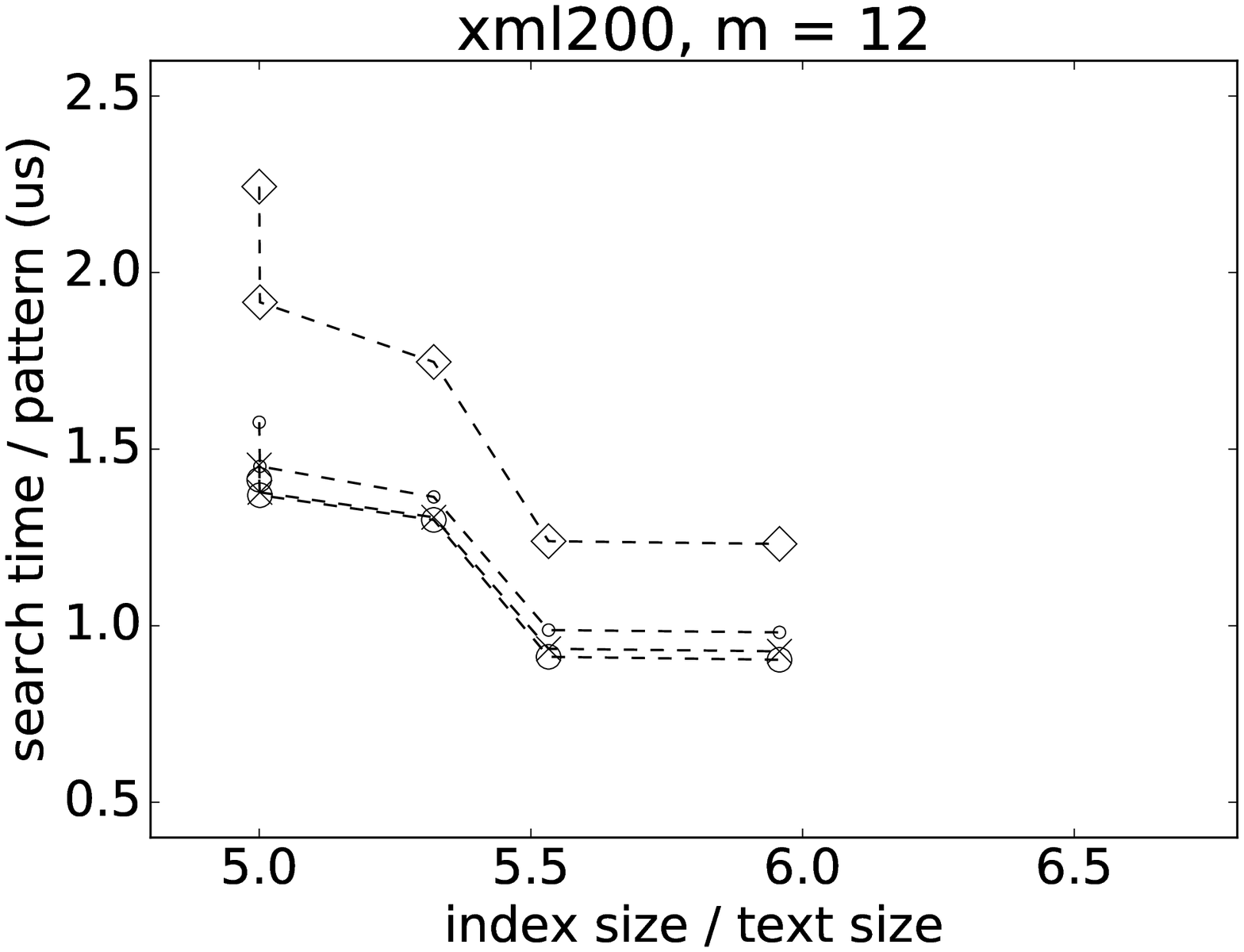}
}
\centerline{
\includegraphics[width=0.42\textwidth,scale=1.0]{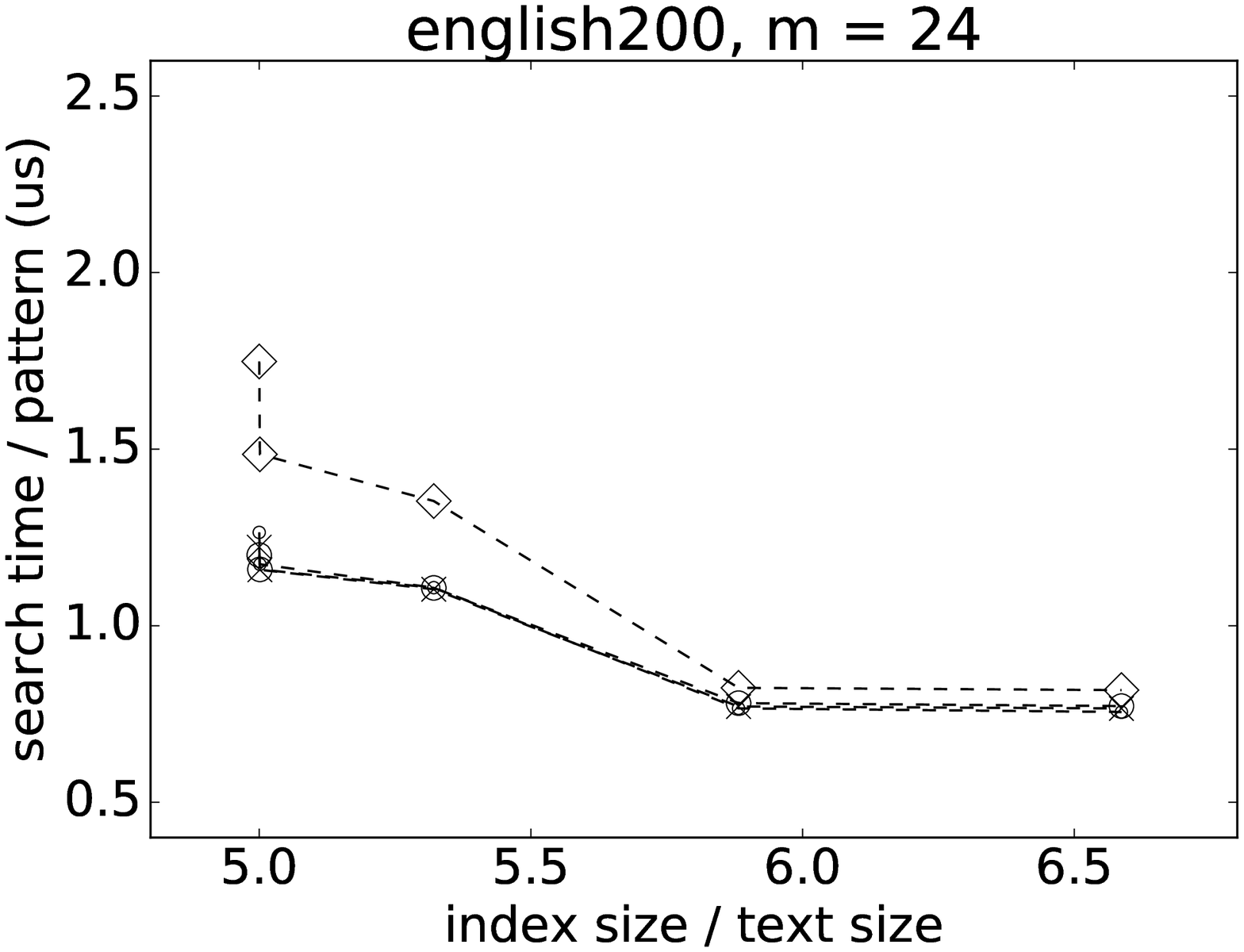}
\includegraphics[width=0.42\textwidth,scale=1.0]{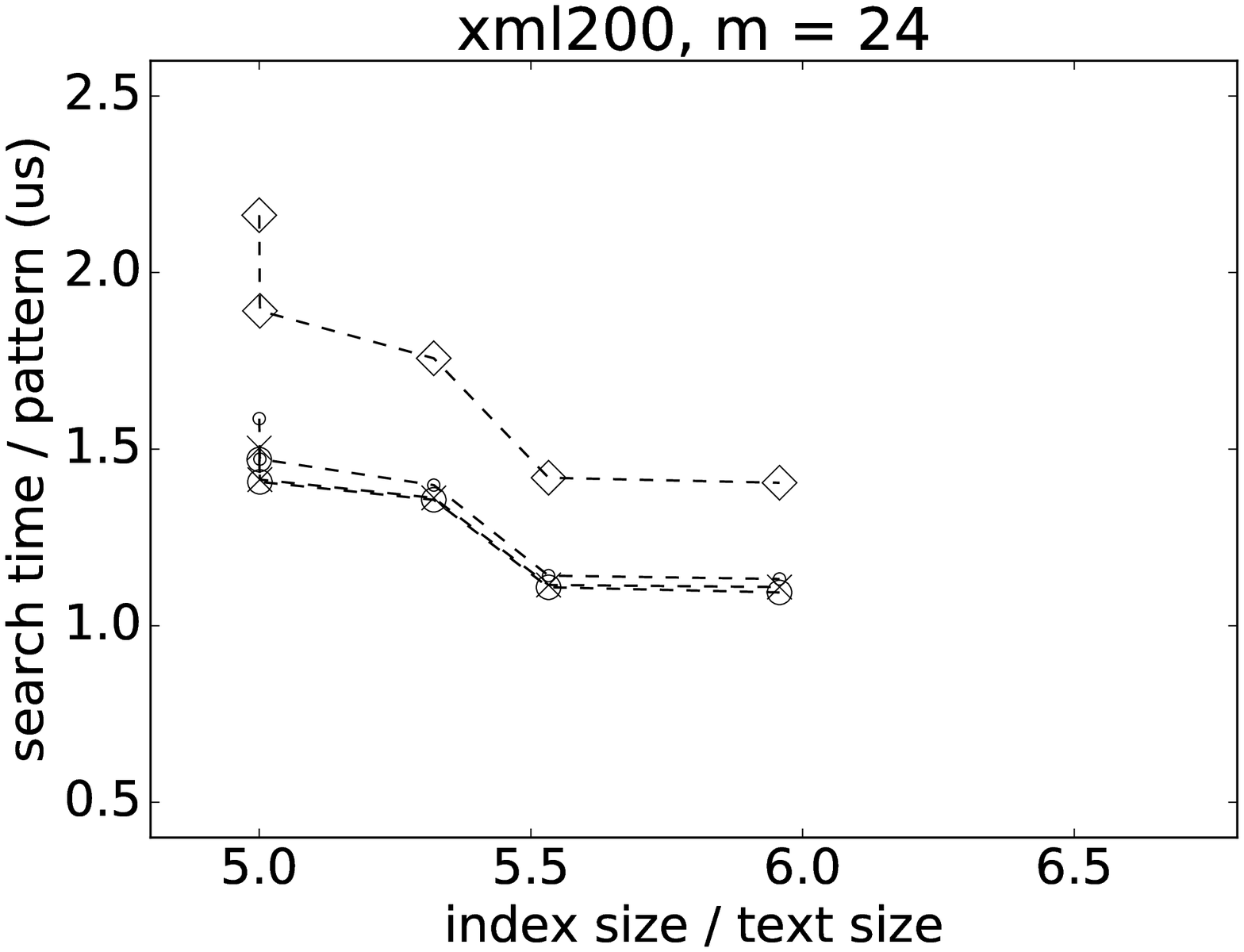}
}
\caption[Results]
{Augmenting the suffix array. The right interval boundary in the standard SA variant 
is found with the doubling technique (-dbl).
The five points in each series correspond to: 
no extra data, LUT2, LUT3, HT with LF=0.9, HT with LF=0.5.}
\label{fig:sa_luts}
\end{figure}

\begin{figure}
\centerline{
\includegraphics[width=0.42\textwidth,scale=1.0]{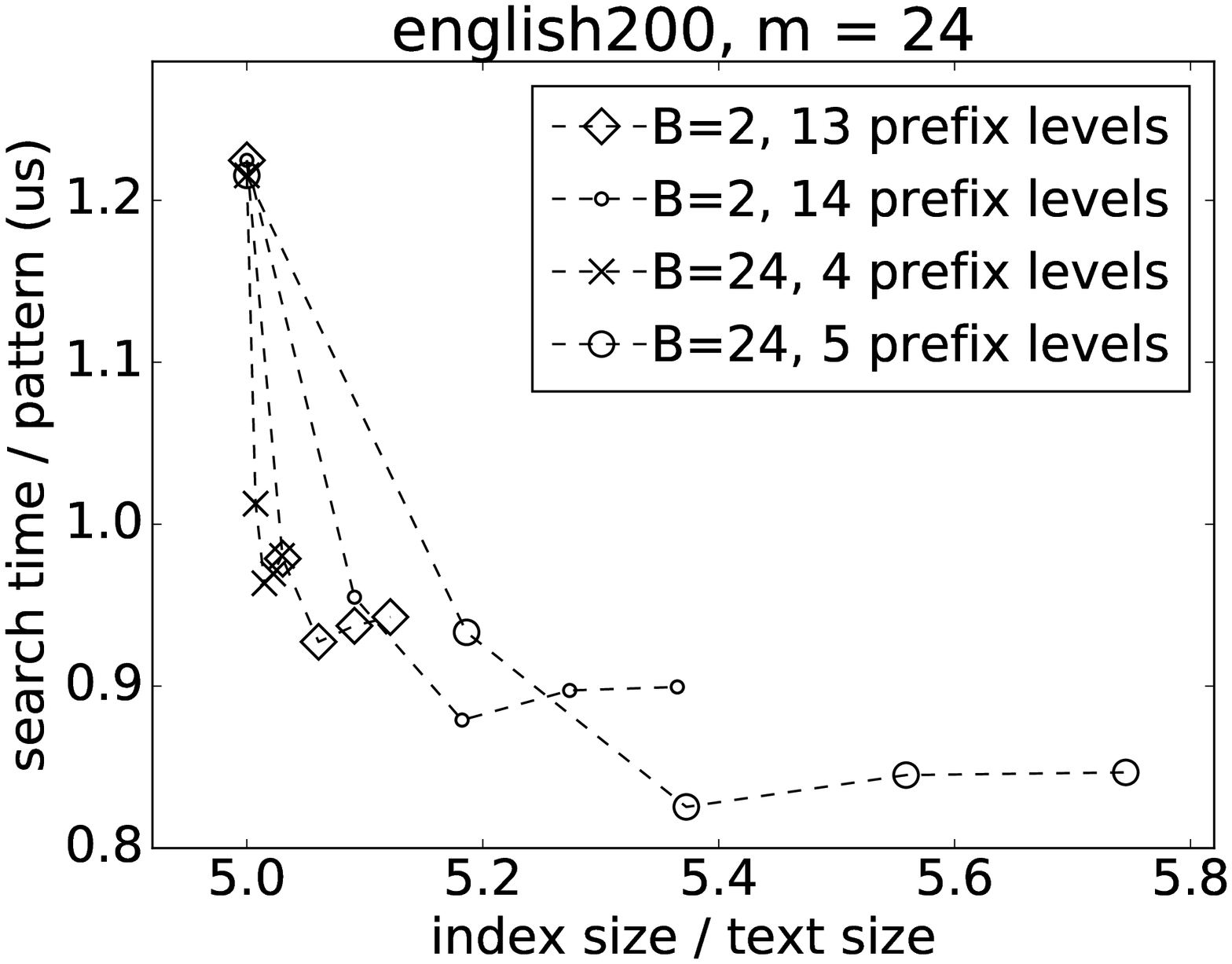}
\includegraphics[width=0.42\textwidth,scale=1.0]{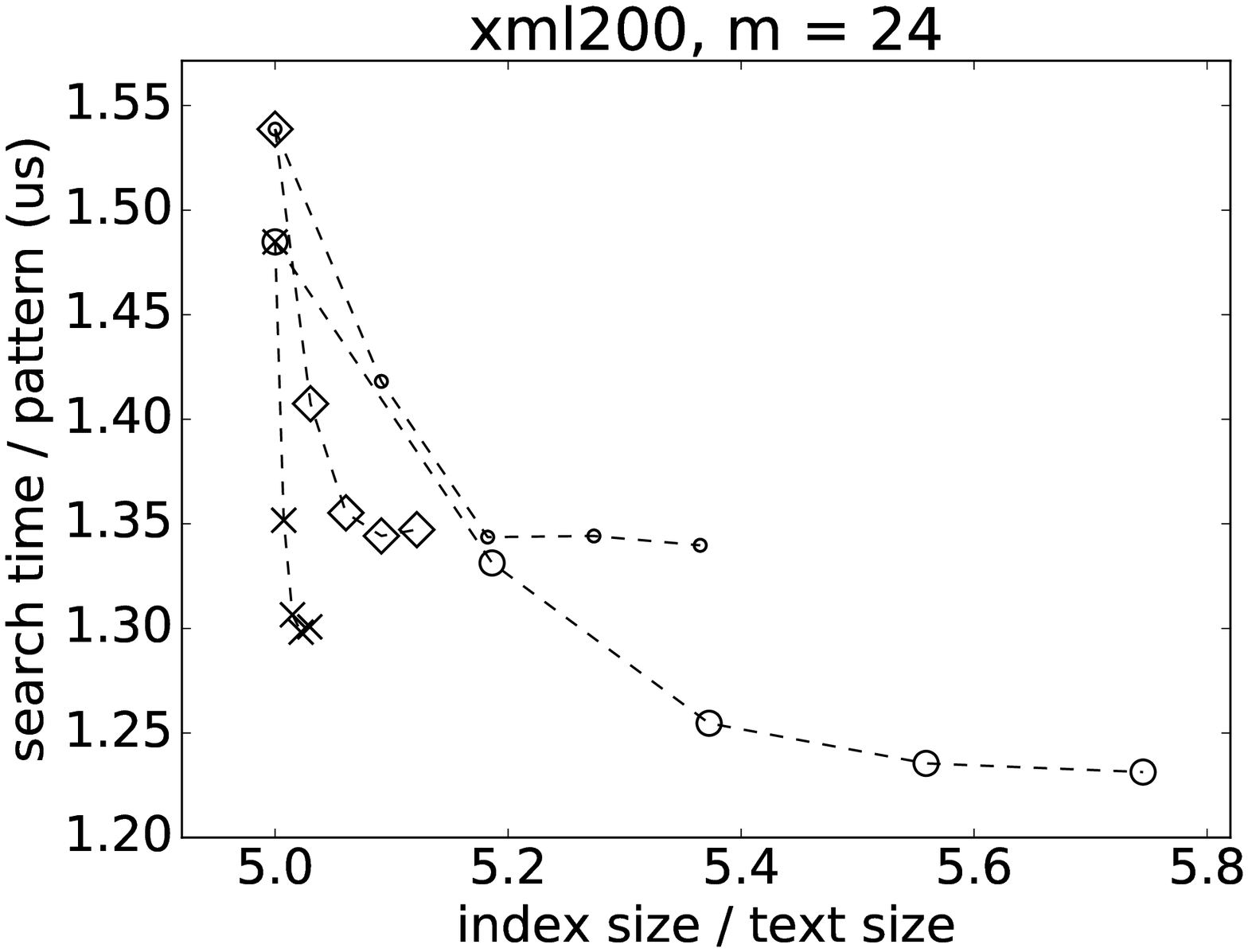}
}
\caption[Results]
{Index sizes and count times for the SA with the B-tree layout, when 
several top levels of the tree store the corresponding suffixes' prefixes 
of length $\{0, 4, 8, 12, 16\}$.} 
%% (The leftmost points in the series correspond to the empty prefix.)}
\label{fig:sa_pl}
\end{figure}

\begin{figure}
\centerline{
\includegraphics[width=0.42\textwidth,scale=1.0]{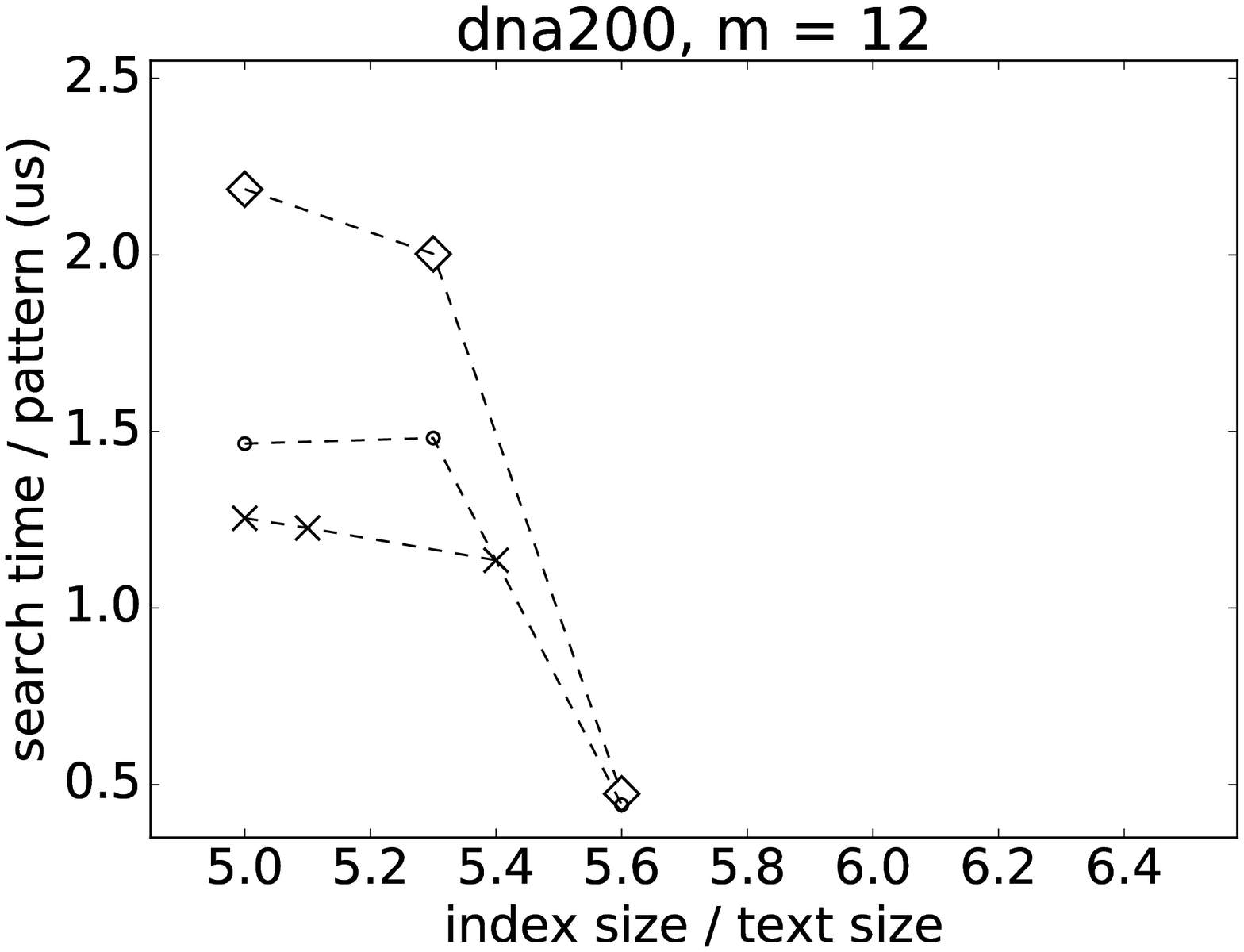}
\includegraphics[width=0.42\textwidth,scale=1.0]{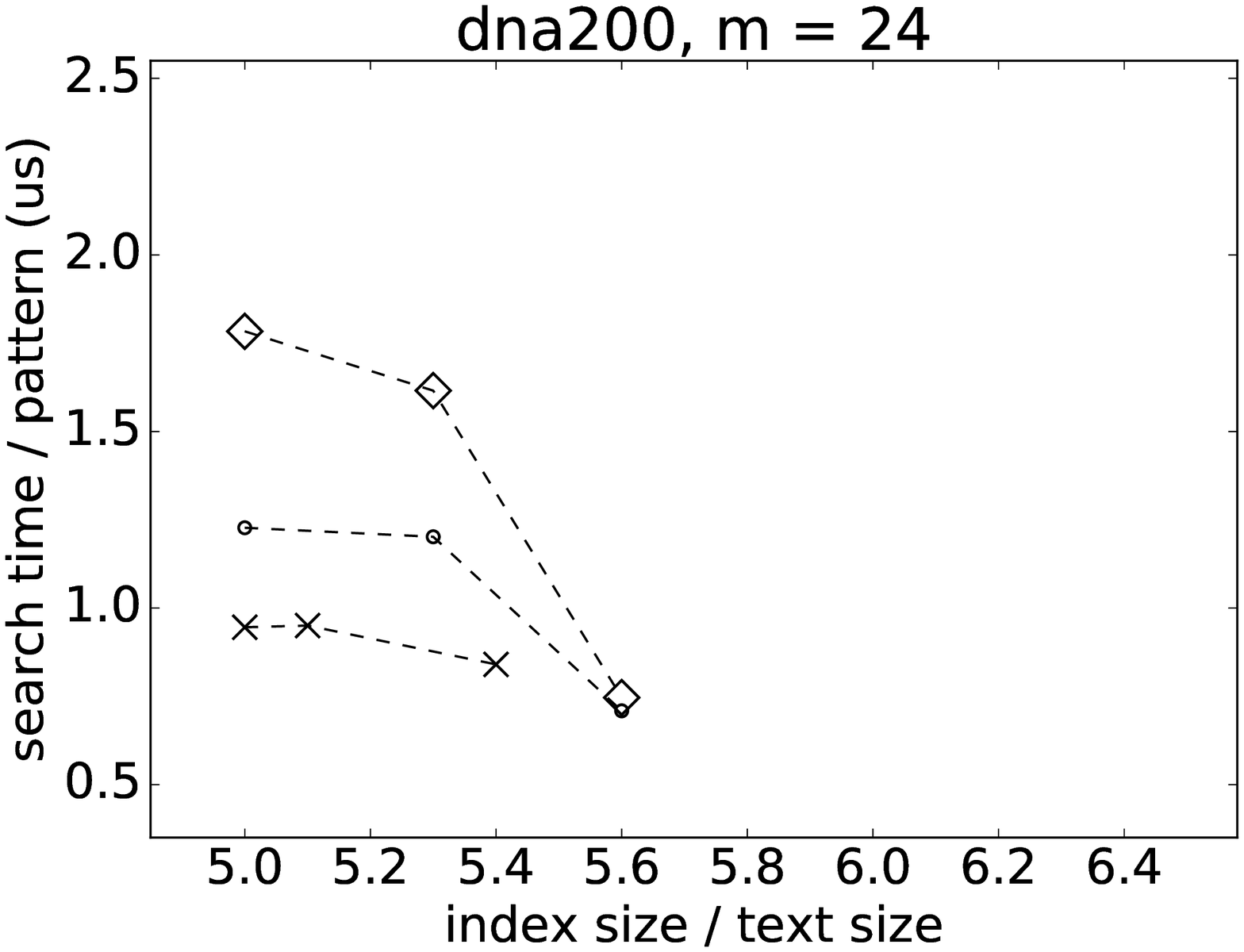}
}
\centerline{
\includegraphics[width=0.42\textwidth,scale=1.0]{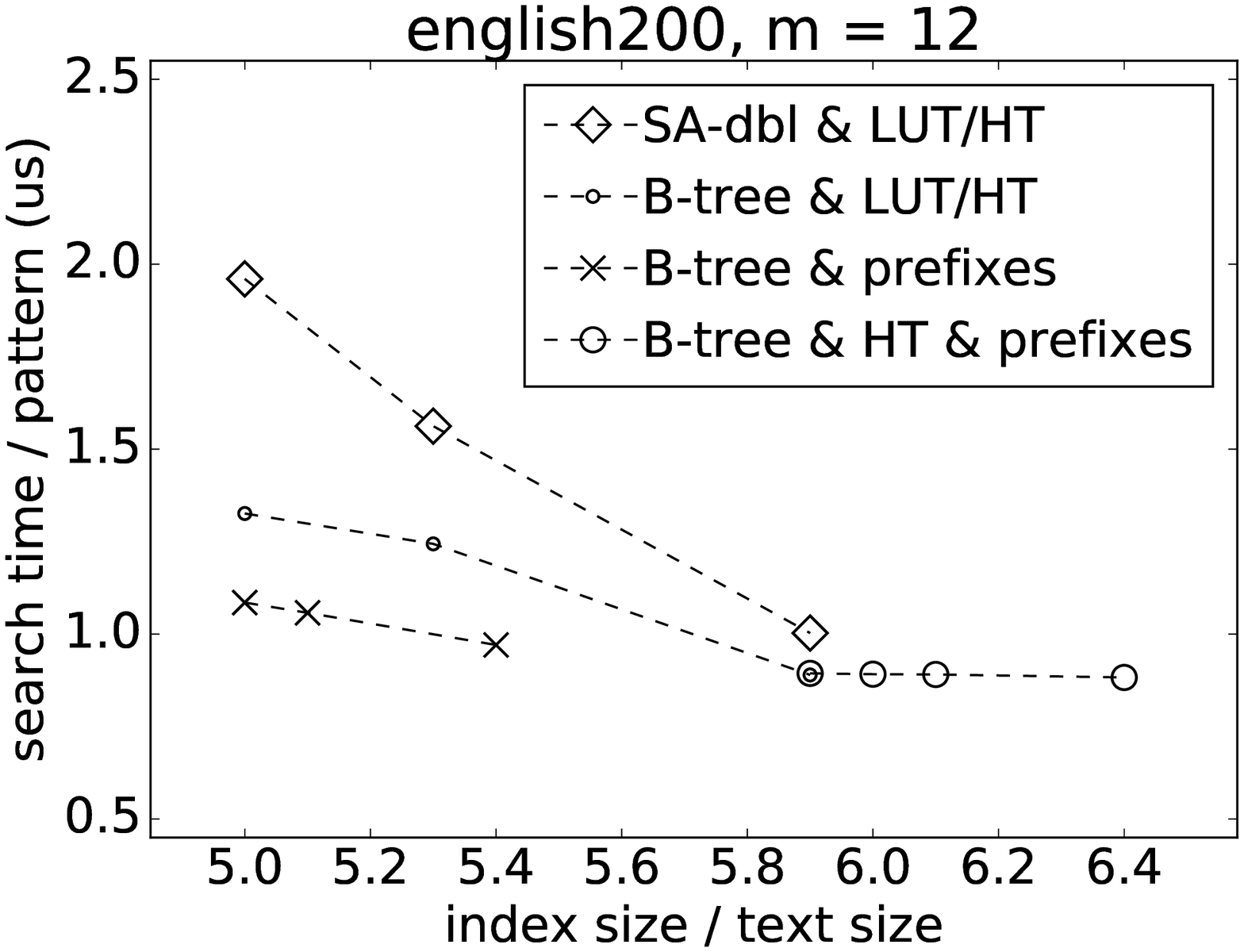}
\includegraphics[width=0.42\textwidth,scale=1.0]{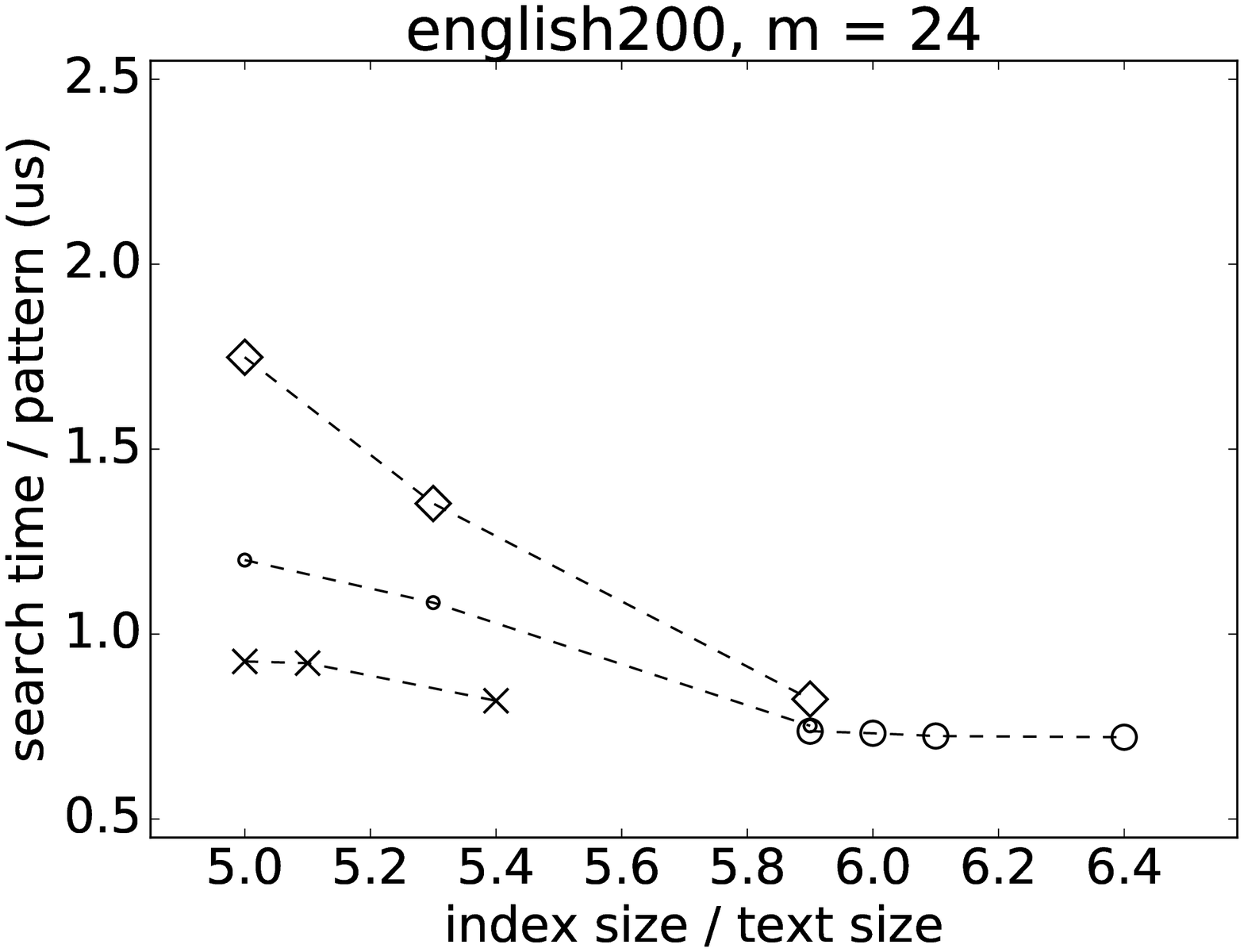}
}
\centerline{
\includegraphics[width=0.42\textwidth,scale=1.0]{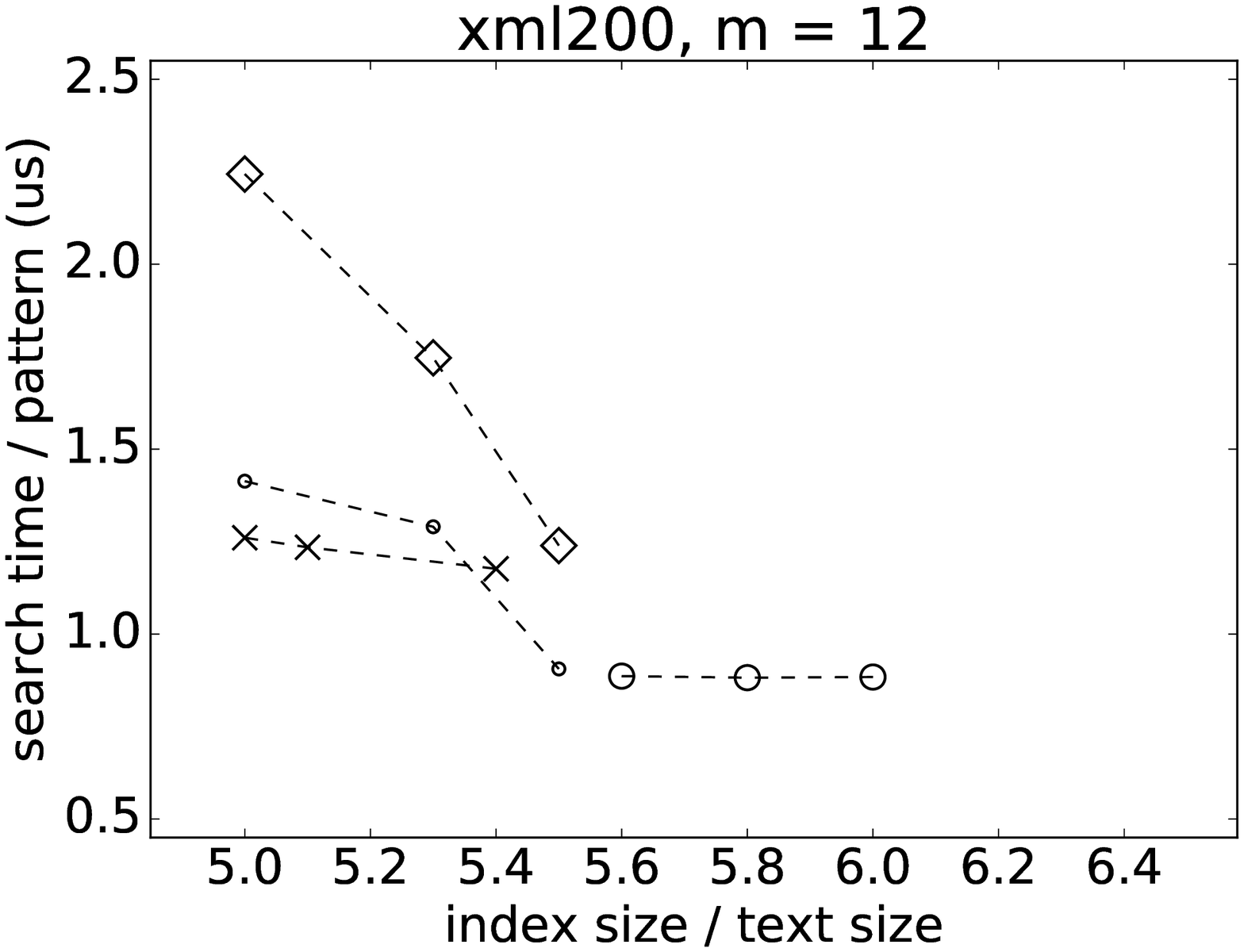}
\includegraphics[width=0.42\textwidth,scale=1.0]{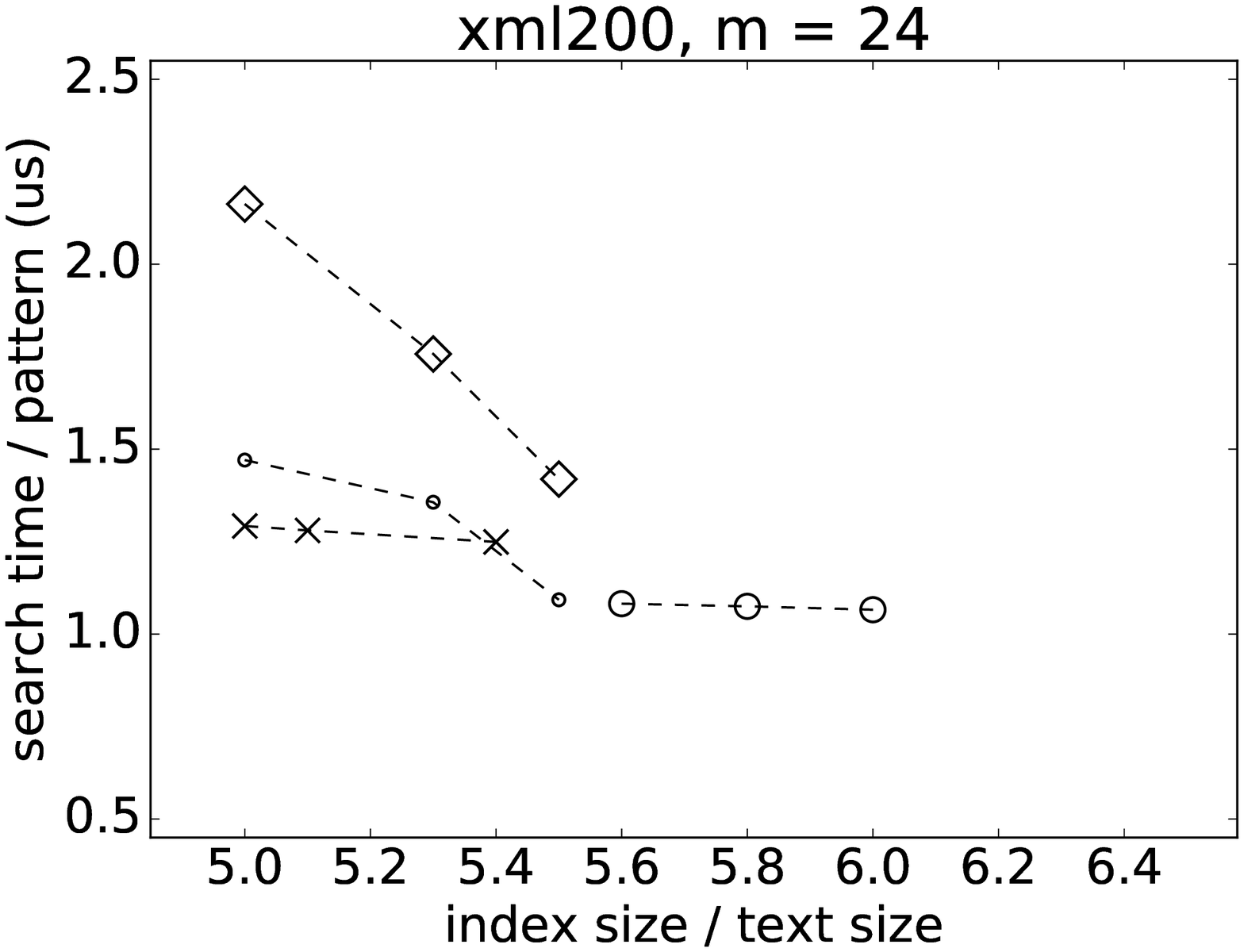}
}
\caption[Results]
{Index sizes and count times for several SA variants with different 
%% (standard vs B-tree) 
layouts and extra data.
Successive points in the series are obtained by changing the 
LUT or hash table component and/or using the prefix copies 
on varying number of levels in the tree.}
%% Successive points in the series stand for time-space tradeoffs.}
\label{fig:sa_btree_lut_pref}
\end{figure}

In Fig.~\ref{fig:sa_luts} we show how augmenting the SA with various structures 
reducing the initial search interval affects the query times and the used space.
%% The lookup table variants (LUT2 and LUT3) work on the byte level 
%% (i.e., we assume $\sigma = 256$) and 
%% The hash table (HT) stores 8-grams from the text and uses xxhash  (\url{https://code.google.com/p/xxhash/}) 
%% as the hash function, with load factor (LF) set to 0.9 or 0.5.
%% LUT2 gives a significant boost with a negligible extra space, 
%% yet it is the extra hash table on 8-grams that excels here 
%% (LF=0.9 is clearly preferred), 
%% speeding up the baseline variant by a factor of 1.5--2.
The hash table (HT), based on the xxhash function 
(\url{https://code.google.com/p/xxhash/}), stores 8-grams from the text 
and was tried with two load factors (LF).
LUT2 gives a significant boost in a tiny space, 
yet it is the hash table (LF=0.9) that excels here, 
%% (LF=0.9 is clearly preferred), 
speeding up the baseline variant by a factor of 1.5--2.

As our SAs with the B-tree layout can be augmented with prefixes 
of the suffixes visited in the first steps 
of the traversal of the tree (i.e., in the top levels), 
we test the impact of the prefix length on the 
performance and space of the resulting index (Fig.~\ref{fig:sa_pl}).
Adding the prefixes gives a noticeable speedup even if they are limited 
in length to 8 characters and are attached to a few tree levels only,
while the space overhead is rather small.
Longer prefixes and more levels help less for a much bigger space penalty.
% Adding short prefixes (of length up to 8) on a few tree levels only gives 
% a noticeable speedup while the space overhead is rather small; 
% longer prefixes and more levels help less for a much bigger space penalty.
%% Adding even short prefixes (of length up to 8) gives a noticeable speedup 
%% while the space overhead is rather small; longer prefixes help little if at all.
Combinations of all ideas presented in the paper are shown 
in Fig.~\ref{fig:sa_btree_lut_pref}, 
where the best option is to combine the B-tree layout with LUT/HT 
(adding prefixes on top of it has a negligible effect).
In total, the speed of the standard SA with the standard right interval 
boundary search was usually improved by a factor exceeding 3
(from 2.6 for xml200 to 3.9 for dna200, for $m = 24$).

%% \section*{Acknowledgement}
\footnotesize
\paragraph{\bf Acknowledgment}

The work was supported by the Polish National Science Centre under the project DEC-2013/09/B/ST6/03117 (the first two and the fourth author).

\normalsize

\bibliographystyle{abbrv}
\bibliography{twist}

\end{document}